\documentclass[12pt]{iopart}

\usepackage{graphicx}
\usepackage{epstopdf}
\usepackage{bm}
\DeclareGraphicsExtensions{.JPG,.eps}

\begin{document}

\title[Post-Thermal Quench Shattered Pellet Injection for small Runaway Electron seed depletion in ITER]{Post-Thermal Quench Shattered Pellet Injection for small Runaway Electron seed depletion in ITER}


\author{E. Nardon$^1$, A. Matsuyama$^2$, D. Hu$^3$, F. Wieschollek$^4$}

\address{$^1$CEA, IRFM, F-13108 Saint-Paul-lez-Durance, France}
\address{$^2$QST, Rokkasho Fusion Institute, Aomori 039-3212, Japan}
\address{$^3$School of Physics, Beihang University, Beijing 100191, China}
\address{$^4$Max Planck Institute for Plasma Physics, Boltzmannstr. 2, 85748 Garching b. M., Germany}

\ead{eric.nardon@cea.fr}
\vspace{10pt}

\begin{abstract} 
The possibility of using Shattered Pellet Injection(s) after the Thermal Quench phase of an ITER disruption in order to deplete Runaway Electron (RE) seeds before they can substantially avalanche is studied. Analytical and numerical estimates of the required injection rate for shards to penetrate into the forming RE beam and stop REs are given. How much material could be assimilated before the Current Quench (CQ) becomes too short is also estimated. It appears that, if Hydrogen pellets were used, the required number of pellets to be injected during the CQ would be prohibitive, at least considering the present design of the ITER Disruption Mitigation System (DMS). For Neon or Argon, the required number of pellets, although large, might be within reach of the ITER DMS, but the assimilated fraction would have to be very small. Other materials may be better suited but would require a modification of the ITER DMS.
\end{abstract}

%
%
%
%
%

\section{Introduction}

Shattered Pellet Injection (SPI) is the reference concept for the ITER Disruption Mitigation System (DMS) \cite{Lehnen_IAEA_2018}\cite{Lehnen_JNM_2015}. One of the main objectives of the DMS is the avoidance of large Runaway Electron (RE) beams \cite{Breizman_2019}\cite{Boozer_PoP_2015}\cite{Martin-Solis_NF_2017}. RE generation comprises primary and secondary generation. There are 4 identified primary (also called `seed') RE generation mechanisms in ITER. The first one, the Dreicer mechanism, is expected to be negligible \cite{Martin-Solis_NF_2017}. The second one, the hot tail mechanism, is difficult to predict. It could be very strong for a hot plasma with a fast Thermal Quench (TQ), as suggested by recent experimental findings \cite{Paz-Soldan_2020}. On the other hand, it has been suggested that hot tail generation may be efficiently reduced by a fast plasma dilution before the TQ using pure Deuterium (or Hydrogen) SPI \cite{Nardon_D2_SPI_2020}. The last two RE seed generation mechanisms are the Tritium $\beta$ decay and the Compton scattering of $\gamma$ rays emitted by the activated wall. These are very small but continuous RE seeds which will be present only during the activated phase of ITER operation. The problem in large tokamaks such as ITER is that even very small RE seeds could give rise to large, multi-MA RE beams because of secondary RE generation by the avalanche effect \cite{Rosenbluth_NF_1997}. This is because the potential number of e-folds in the avalanche is proportional to the poloidal flux contained in the plasma and thus to the plasma current, leading to an avalanche gain $G_{avalanche} \simeq \exp(2.5\times I_p[MA]) \simeq 1.9 \times 10^{16}$ for a 15 MA ITER plasma, according to \cite{Hender_NF_2007}. More recent work suggests that the avalanche gain could be far greater still in the presence of partially ionized impurities \cite{Hesslow_NF_2019}. Thus, it appears that RE seeds may be small in ITER, but still could lead to dangerous RE beams because of the very large avalanche gain.

The present paper discusses the following question: is it possible to deplete RE seeds before they have substantially avalanched, by injecting pellets, possibly shattered, counting on the stopping power of the shards/pellets (in the following, we will use only the term `shards' for concision)? REs will typically travel across shards. Each time they will do so, they will lose a certain amount of energy. If this energy dissipation is fast enough compared to the acceleration by the loop voltage and the population growth from the avalanche, the formation of large RE beams might be avoided. But clearly, since RE seeds are continuously produced, a continuous injection or repeated discrete injections would be required.

The aim of this paper is to provide rough estimates to assess the feasibility of such a scheme. We will first discuss in Section \ref{sec:design_choices} the `degrees of freedom' in the design of the scheme, i.e. the different types of injections that can be envisaged. Then, we shall analyze the 3 main conditions which are required for the scheme to work. The first one is that shards should be able to stop RE seeds. This will be addressed analytically in Section \ref{sec:analytic_estimates} and numerically in Section \ref{sec:simulations}. The second condition is that shards should be able to penetrate the forming RE beam, which will be discussed in Section \ref{sec:penetration}. The third condition, addressed in Section \ref{sec:CQ_timescale}, is that the material possibly assimilated by the plasma as a result of these injections should not make the Current Quench (CQ) faster than tolerable. Finally, Section \ref{sec:summary} will summarize results and discuss priorities for future work. 

\section{Degrees of freedom}
\label{sec:design_choices}

\begin{figure}[ht]
	\centering
	\includegraphics[width=60mm]{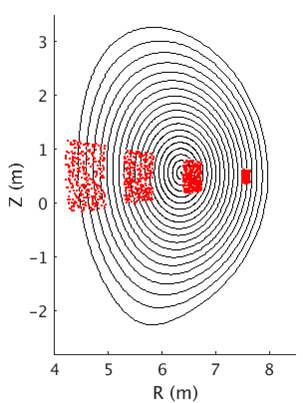}
	\includegraphics[width=60mm]{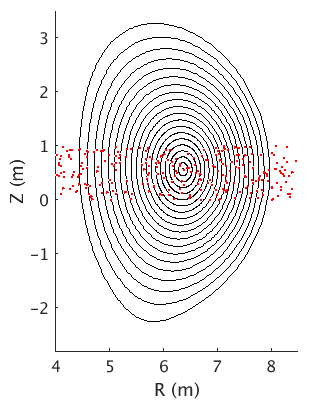}
	\caption{Schematic plots illustrating the concept of repeated (left) or continuous (right) injection(s).}
	\label{fig:repeated_continuous}
\end{figure}

Since RE seed production is a continuous process, the injection scheme should be based either on repeated injections or on a continuous injection, like shown schematically in Fig. \ref{fig:repeated_continuous}. In this figure, the repeated injections (left plot) correspond to successive SPI from the Low Field Side (LFS) midplane, with a certain horizontal and vertical velocity spread of the shards. Such repeated injections are in principle possible with the present design of the ITER DMS. The continuous injection (right plot), on the other hand, is more conceptual and harder to approach with the present ITER DMS design. It consists in a steady beam of shards, also coming from the LFS. All shards are assumed to have the same, purely horizontal, velocity. Degrees of freedom for both types of injection are the time-averaged solid mass flux and the averaged velocity of the shards. In the case of repeated injections, other degrees of freedom are the time delay between injections and the shards velocity spread. 

\begin{figure}[ht]
	\centering
	\includegraphics[width=30mm]{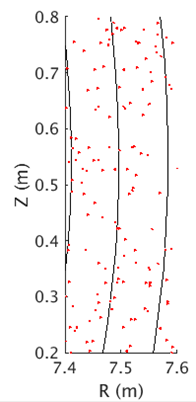}
	\includegraphics[width=30mm]{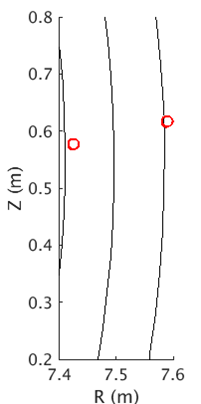}
	\caption{Schematic illustratiions of the `many small shards' (left) and `a few large shards' (right) cases.}
	\label{fig:small_large_shards}
\end{figure}

Another important degree of freedom, as we shall see below, for both the repeated and continuous injection, is the size of the shards. Two limiting cases can be distinguished: the case of `many small shards' and the case of `a few large shards', as illustrated in Fig. \ref{fig:small_large_shards}. The key difference between these two cases is that in the former, there is at any instant a number of shards intercepting a given flux surface (during the passage of a shard cloud across that surface), while in the latter, there are temporal gaps between shards passages. Note that the `few large shards' case should not necessarily be based on SPI but could also correspond to repeated injections of non-shattered pellets.

The last important degree of freedom is the pellet material. In this paper, we will consider pure Hydrogen, Neon or Argon pellets since these are the species currently envisaged for the ITER DMS, but the results presented could easily be generalized to other species. The key trade-off is between stopping power and the risk to make the CQ too short, both being much smaller for Hydrogen than for Neon or Argon, as we shall see below.

\section{Can shards stop Runaway Electrons? Analytical estimates}
\label{sec:analytic_estimates}

Let us begin with analytical estimates to assess the possibility to stop REs with shards, assuming the latter penetrate the forming RE beam. To simplify expressions, shards are assimilated to cubes of edge length $2 r_s$, with the same $r_s$ for all shards. REs are typically energetic enough to travel across a shard. When they do so, they lose an energy $\Delta E = 2 r_s p$, where $p$ is the stopping power of the shard (assuming, for simplicity, that the RE crosses the shard in a direction parallel to an edge). The averaged energy lost by a RE per toroidal turn in this process is:

\begin{equation}
	E_{loss/turn} = N_{shard/turn} \Delta E
\end{equation}

where $N_{shard/turn}$ is the average number of shards encountered per turn. In order to estimate $N_{shard/turn}$, in what follows we will assimilate RE drift surfaces to flux surfaces and assume that the poloidal position of a RE aftera certain number of toroidal turns is a random variable with a homogenous probability density over $[0,2\pi]$ (this assumption will be justified in Section \ref{sec:rational_q}). 

The energy gained by a RE from the loop voltage per toroidal turn is: 

\begin{equation}
           E_{gain/turn} = e V_{loop}.
\end{equation}

We will use the value $V_{loop} = 1.1$ kV, which is expected at the beginning of a CQ of exponential time constant 66 ms (corresponding to a duration of 150 ms in the `usual' definition which consists in taking the exponential decay time between 80 and 20\% of the initial plasma current and dividing it by 0.6) for an initial poloidal flux of 75 V.s, which is a typical value for an ITER 15 MA plasma \cite{Boozer_PPCF_2019}. We chose here a CQ timescale in the upper range of what is considered tolerable, which is favorable for the discussed RE depletion scheme. On the other hand, we take $V_{loop}$ at the beginning of the CQ, conservatively ignoring the fact that $V_{loop}$ decays during the course of the CQ.

A first condition for a successful depletion of the RE seed population is: 

\begin{equation}
          \frac{E_{gain/turn}}{E_{loss/turn}} < 1.
          \label{eq:Egt_ov_Elt}
\end{equation}

This condition is sufficient in the case of a continuous injection of many small shards. On the other hand, for repeated injections separated by gaps or for the case of a few large shards, REs may re-accelerate during the time intervals when no shards are passing (the acceleration time of an electron from rest up to 10 MeV, neglecting any drag, is $\simeq 1$ ms in ITER with $V_{loop} = 1.1$ kV). In these cases, another necessary condition for success is thus:

\begin{equation}
          \frac{t_{stop}}{t_{pass}} < 1
\end{equation}

where $t_{stop}$ is the time it takes to stop a RE and $t_{pass}$ is the time during which the shard cloud (in the case of repeated injections with many small shards) or an individual shard (in the case of a few large shards) passes across the surface where the RE is located.
 
 We shall now provide estimates for $E_{gain/turn}/E_{loss/turn}$ and $t_{stop}/t_{pass}$ for the case of many small shards in Section \ref{many_small_shards} and for the case of a few large shards in Section \ref{few_large_shards}.

\subsection{Case of many small shards}
\label{many_small_shards}

In the case of many small shards, the number of shards which intersect a given flux surface is $N_{s@\psi} \simeq 2 r_s dN_s/dr_\psi$, where $dN_s/dr_\psi$ is the shard number density with respect to the minor radial coordinate $r_\psi$, and the $2 r_s$ factor is an estimate of the extension of shards along the minor radius.

The fraction of the poloidal perimeter of the flux surface which is intercepted by a given shard is $f_\theta \simeq 2r_s/(2\pi r_\psi)$, assuming flux surfaces have a circular cross-section. It can be remarked that, due to the $1/r_\psi$ factor, shards will brake REs more easily in the core than at the edge. The average number of shards a RE encounters per toroidal turn is then:

\begin{equation}
	N_{shard/turn} = N_{s@\psi} f_\theta \simeq \frac{2 r_s^2}{\pi r_\psi} \frac{dN_s}{dr_\psi},
\end{equation}

resulting in an average energy loss per turn:

\begin{equation}
	E_{loss/turn} \simeq \frac{4 r_s^3 p}{\pi r_\psi} \frac{dN_s}{dr_\psi}
	\label{eq:Elt_many_small_1}
\end{equation}

It thus appears that $E_{loss/turn} \propto d(N_s r_s^3 )/dr_\psi \propto d(N_s V_s)/dr_\psi$, where $V_s$ is the shard volume. In other terms, $E_{loss/turn}$ is proportional to the radial derivative of the total volume occupied by the shard population. This can be understood when considering the problem from the point of view of shards instead of electrons. Indeed, shards are permanently being travelled across by a RE flux of density $cn_{RE}$, where $c$ is the speed of light and $n_{RE}$ is the RE density (assumed homogeneous). This corresponds to a power dissipation in each shard equal to $c n_{RE} p V_s$. The radial derivative of the power dissipated in shards is thus $c n_{RE} p V_s dN_s/dr_\psi$. The power lost by each RE is approximately $E_{loss/turn}c/(2\pi R)$, where $R$ is the major radius. Thus the radial derivative of the power lost by all REs is $E_{loss/turn}c/(2\pi R)dN_{RE}/dr_\psi$, with $dN_{RE}/dr_\psi = n_{RE} \times 2\pi r_\psi \times 2 \pi R$. Equating the two preceding expressions for the radial derivative of the power dissipated in shards, one finds:

\begin{equation}
	E_{loss/turn}  \simeq \frac{V_s p}{2 \pi r_\psi } \frac{dN_s}{dr_\psi}.
	\label{eq:E_loss_v2}
\end{equation}

Considering that, to obtain Eq. \ref{eq:Elt_many_small_1}, we had assimilated shards to cubes of edge length $2 r_s$ and thus of volume $8 r_s^3$, Eqs. \ref{eq:Elt_many_small_1} and \ref{eq:E_loss_v2} are consistent with each other. However, Eq. \ref{eq:E_loss_v2} is more general.

An important comment here is that $E_{loss/turn}$ does not depend on the shard radius $r_s$, i.e. the shard size distribution does not affect $E_{loss/turn}$ (as long as the `many small shards' regime is valid).

Let us make first numerical estimations. The minimal value (with respect to the electron energy) of the ESTAR stopping power is $3.8$ MeV.cm$^2$/g for Hydrogen, $1.6$ MeV.cm$^2$/g for Neon and $1.4$ MeV.cm$^2$/g for Argon \cite{ESTAR} which, multiplied by the respective solid density of these materials, corresponds to $38$ MeV/m for Hydrogen and $230$ MeV/m for both Neon and Argon. Condition \ref{eq:Egt_ov_Elt} thus corresponds to $V_s dN_s/dr_\psi > 5.3 V_{\small{\textrm{28mm}}}$/m for Hydrogen and $V_s dN_s/dr_\psi > 0.87 V_{\small{\textrm{28mm}}}$/m, where $V_{\small{\textrm{28mm}}}$ is the volume of a large ITER DMS pellet under the present design, i.e. of a cylinder a diameter 28 mm and length 56 mm ($V_{\small{\textrm{28mm}}} = 3.4 \times 10^{-5}$ m$^3$). We will refer to these pellets as `28 mm pellets' in the following. This means that the volume of shard material should correspond to more than 5.3 (resp. 0.87) 28 mm pellets per meter.

We shall now specialize the above expression for $E_{loss/turn}$ to the case of a shard cloud from a single SPI. In this case, $dN_s/dr_\psi = N_s/L_{cloud}$, where $N_s$ is the number of shards in the cloud and $L_{cloud}$ is the radial extension of the cloud. The latter is related to the velocity spread of the shards $\Delta v_s$ in the following way: $L_{cloud} = l\Delta v_s /\langle v_s \rangle$, where $\langle v_s \rangle$ is the average shard velocity and $l = r_{s0} - r_\psi$ is the distance shards need to travel to reach the flux surface under consideration from their initial position $r=r_{s0}$. Thus:

\begin{equation}
	E_{loss/turn} \simeq \frac{V_s p N_s \langle v_s \rangle}{2\pi r_\psi (r_{s0}-r_\psi) \Delta v_s}.                         
\end{equation}

\begin{figure}[ht]
	\centering
	\includegraphics[width=80mm]{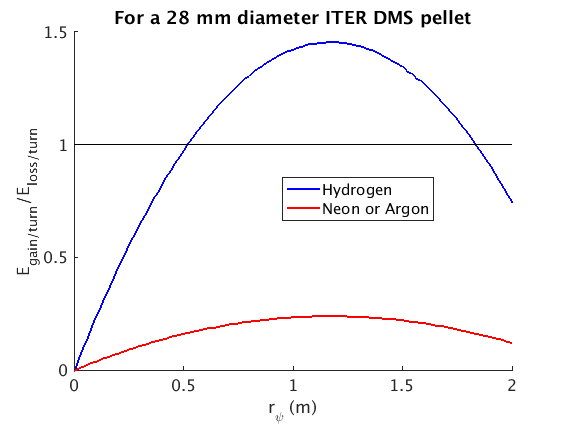}
	\caption{$E_{gain/turn}/E_{loss/turn}$ ratio as a function of the flux surface minor radius $r_\psi$, for $\Delta v_s /\langle v_s \rangle = 0.2$ and for a 28 mm Hydrogen, Neon or Argon pellet shattered into many small shards, taking the minimal (with respect to the electron energy) ESTAR stopping power and considering that $V_{loop}=1100$ V.}
	\label{fig:Egain_over_Eloss}
\end{figure}

In the present ITER DMS design, $r_{s0} \simeq a + 0.35$ m, where $a \simeq 2$ m is the plasma minor radius \cite{Lehnen_private}. Fig. \ref{fig:Egain_over_Eloss} shows $E_{gain/turn}/E_{loss/turn}$ as a function of $r_\psi$, for $\Delta v_s /\langle v_s \rangle = 0.2$ and for a Hydrogen, Neon or Argon 28 mm pellet. It can be seen that a single Hydrogen pellet is not sufficient to brake REs everywhere in the plasma, although it is sufficient in the very core and very edge, but two synchronized pellets should be sufficient. On the other hand, a single Neon or Argon pellet should easily be able to brake REs throughout the plasma, and one fourth of a pellet should be marginally sufficient.

As discussed above, for the case of repeated SPI with gaps between shard clouds, we should also estimate $t_{stop}/t_{pass}$. The stopping time is: 

\begin{equation}
         t_{stop} \simeq \frac{2\pi R}{c} \frac{E_{RE}^0}{E_{loss/turn}-E_{gain/turn}} 
\end{equation}

where $E_{RE}^0$ is the initial RE energy. The time it takes for the shard cloud to pass across a flux surface is:

\begin{equation}
         t_{pass} = \frac{L_{cloud}}{\langle v_s \rangle} = (r_{s0}-r_\psi) \frac{\Delta v_s}{\langle v_s \rangle^2}.
\end{equation}

We thus have:

\begin{equation}
        \frac{t_{stop}}{t_{pass}} = \frac{2\pi R}{c} \frac{E_{RE}^0}{E_{loss/turn}-E_{gain/turn}}  \frac{1}{r_{s0}-r_\psi} \frac{\langle v_s \rangle^2}{\Delta v_s}.
\end{equation}

In the limit $E_{loss/turn} \gg E_{gain/turn}$, this simplifies to:

\begin{equation}
        \frac{t_{stop}}{t_{pass}} \simeq \frac{4\pi^2 R r_\psi E_{RE}^0  \langle v_s \rangle}{c N_s V_s p}.
        \label{eq:t_brake_t_cloud}
\end{equation}

Not suprisingly, $t_{stop}/t_{pass}$ is proportional to $\langle v_s \rangle$, meaning that slower shards are beneficial for stopping REs. On the other hand, $t_{stop}/t_{pass}$ is independent of $\Delta v_s$. This is because both $t_{stop}$ and $t_{pass}$ are proportional to $1/\Delta v_s$: a cloud with less velocity spread is more `packed together' and thus brakes REs faster due to its larger shard density, but it also passes faster across the surface, and the two effects compensate each other.

\begin{figure}[ht]
	\centering
	\includegraphics[width=70mm]{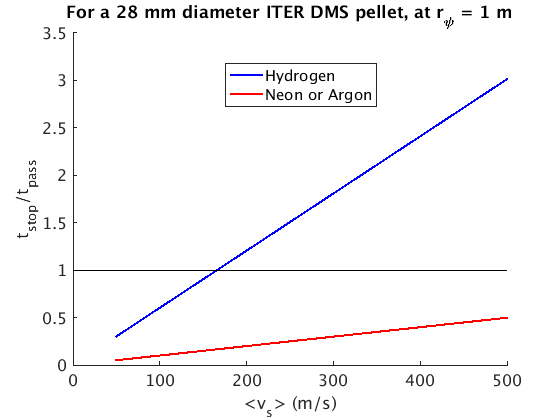}	
	\includegraphics[width=70mm]{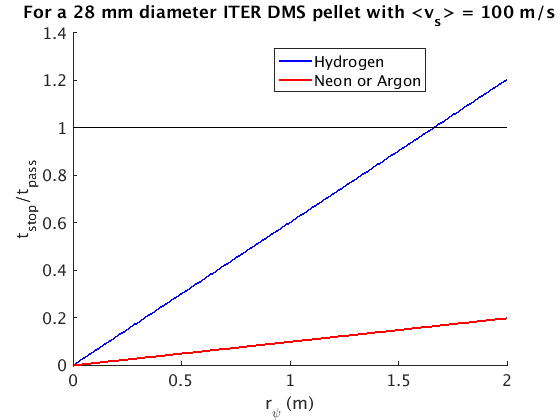}
	\caption{$t_{stop}/t_{pass}$ (in the limit where $E_{loss/turn} \gg E_{gain/turn}$) as a function of $\langle v_s \rangle$ at $ r_\psi = 1$ m (i.e. mid-radius) (left) and as a function of $ r_\psi$ for $\langle v_s \rangle = 100$ m/s (right) for a single 28 mm Hydrogen, Neon or Argon ITER DMS pellet shattered into many small shards. The black horizontal line indicates $t_{stop}/t_{pass} = 1$, i.e. the limit above which REs are not fully stopped.}
	\label{fig:tbrake_over_tcloud}
\end{figure}

Fig. \ref{fig:tbrake_over_tcloud} shows $t_{stop}/t_{pass}$ as a function of $\langle v_s \rangle$ at $ r_\psi = 1$ m (i.e. mid-radius) (left) and as a function of $ r_\psi$ for $\langle v_s \rangle = 100$ m/s (right) for a single 28 mm Hydrogen, Neon or Argon pellet. It appears that a Neon or Argon pellet easily fulfills $t_{stop}/t_{pass} < 1$, while a Hydrogen pellet needs to be slow enough and may be efficient only in the core region (furthermore, the assumption $E_{loss/turn} \gg E_{gain/turn}$ does not hold for such a pellet as we have seen above, so Fig. \ref{fig:tbrake_over_tcloud} is over-optimistic).

\subsection{Case of a few large shards}
\label{few_large_shards}

Let us now consider the case of a few large shards. As in the previous section, we shall begin by assimilating shards to cubes of side length $2r_s$ for simplicity. When a shard intersects a given flux surface, it occupies a fraction $\simeq 2 r_s / (2 \pi r_\psi)$ of its poloidal perimeter, which corresponds to the probability that a RE located on that surface passes through the shard per toroidal turn. When multiplied by $\Delta E = 2 r_s p$, this results in:

\begin{equation}
	E_{loss/turn} \simeq \frac{4 r_s^2p}{2 \pi r_\psi}.
	\label{eq:Elt_few_large_1}
\end{equation}

If we now consider the more realistic case of spherical shards, the averaged value of $E_{loss/turn}$ over the passage of the shard across the flux surface can be calculated by integrating over the shard passage:

\begin{equation}
	E_{loss/turn} \simeq \int_{x=0}^{2r_s} \frac{4(r_s^2-x^2)p}{2\pi r_\psi}\frac{dx}{2r_s} = \frac{V_s p}{2 \pi r_\psi} \frac{2}{\pi r_s},
	\label{eq:Elt_few_large_2}
\end{equation}

providing a more general expression for $E_{loss/turn}$ which is consistent with Eq. \ref{eq:Elt_few_large_1}. Eq. \ref{eq:Elt_few_large_2} is also consistent with the expression obtained for the case of many small shards, Eq. \ref{eq:E_loss_v2}, replacing $dN_s/dr_\psi$ by $ 2/(\pi r_s)$.

The condition $E_{gain/turn}/E_{loss/turn} < 1$ can then be written: 

\begin{equation}
	r_s > \sqrt{\frac{3 \pi r_\psi e V_{loop}}{4 p}},
	\label{eq:rs_few_large_shards}
\end{equation}

which translates to $r_s > 8.3$ mm for Hydrogen and $r_s > 3.4$ mm for Neon or Argon, for $r_\psi = 1$ m.

The time it takes for a spherical shard to pass across a flux surface is $t_{pass} = 2 r_s / v_s$, where $v_s$ is the shard radial velocity. We thus have, making use of Eq. \ref{eq:Elt_few_large_2} and assuming $E_{loss/turn} \gg E_{gain/turn}$:

\begin{equation}
        \frac{t_{stop}}{t_{pass}} \simeq  \frac{\pi^3 RE_{RE}^0 r_\psi  v_s }{cV_s p},
        \label{eq:t_stop_t_cloud_single_shard}
\end{equation}

which translates to $r_s > 16$ mm for Hydrogen and $r_s > 8.6$ mm for Neon or Argon, assuming $v_s = 100$ m/s. These shard radii are roughly twice larger than those found from the condition $E_{gain/turn}/E_{loss/turn} < 1$, thus justifying the assumption $E_{loss/turn} \gg E_{gain/turn}$ \textit{a posteriori} (note that $E_{loss/turn} \propto r_s^2$). It therefore appears that in the `few large shards' regime, shards indeed need to be quite large. Assuming that shards result from the shattering of a 28 mm pellet, the requirement $r_s > 16$ mm (resp. $r_s > 8.6$ mm) corresponds to no more than $\simeq 2$ (resp. $\simeq 13$) shards per pellet, which appears difficult if not impossible. Repeated injections of smaller non-shattered pellets are an alternative.

\section{Can shards stop Runaway Electrons? Numerical simulations}
\label{sec:simulations}

The above numerical estimates are useful as a first approach to the problem, but they are based on several simplifying assumptions. As a second step in our study, we will now describe more realistic, although still largely simplified, numerical simulations.

\subsection{Model description}

For simplicity, REs are assumed to follow field lines. It is assumed that all shards are spheres with the same radius $r_s$, which remains constant in time. The model evolves by discrete steps, each step corresponding to one toroidal turn made by the REs. At each step, time is incremented by $2\pi R/v_{RE}$, where $v_{RE}$ is the RE velocity from the previous step (each RE has its own time), and the poloidal angle of the electron is incremented by $2\pi/q$, where $q$ is the safety factor. Then, the position $(R_{RE},Z_{RE})$ of the RE in Cartesian coordinates is calculated and the RE energy is incremented by:

\begin{equation}
	\Delta E_{RE} = e V_{loop} - p  \sum_{i_s=1}^{N_s} d_i 
\end{equation}

where $d_i \equiv 2 \sqrt{\max(r_s^2-(R_{RE}-R_s(i_s,t))^2-(Z_{RE}-Z_s(i_s,t))^2,0)}$ is the distance travelled across each shard, with $(R_s(i_s,t),Z_s(i_s,t))$ the position of shard $i_s$ at time $t$. However, the RE energy is not allowed to overcome 10 MeV, which is a simplified way to account for energy limiting processes such as synchrotron radiation. The new RE velocity is then calculated from its updated energy by $v_{RE} = c \times \sqrt{1-1/\gamma^2}$, with $\gamma = E_{RE}/(m_ec^2)$, $m_e$ being the electron rest mass. 

\begin{figure}[ht]
	\centering
	\includegraphics[trim=100 0 100 0, clip, width=50mm]{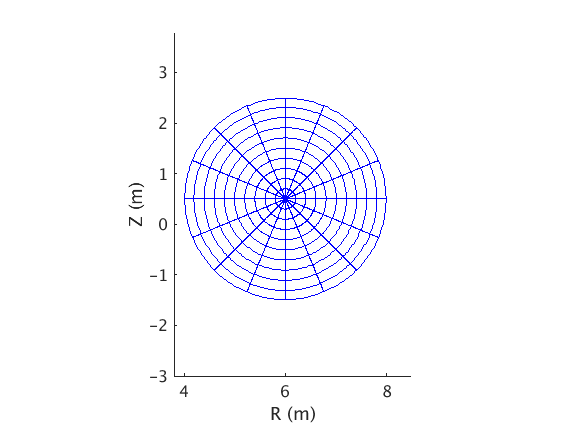}
	\includegraphics[trim=100 0 100 0, clip, width=50mm]{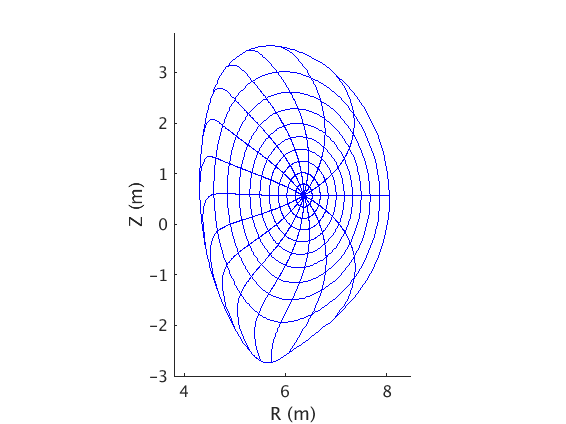}	
	\caption{Geometry of the circular (left) and realistic (right) equilibria used for numerical simulations. Lines show iso-contours of the poloidal flux and of the straight field line poloidal angle (taking the geometric angle as the toroidal angle).}
	\label{fig:equil_geometry}
\end{figure}

We consider two types of geometries. The first one has a circular cross-section with the poloidal angle in straight field line coordinates $\theta^*$ being the geometrical angle (the toroidal angle also being the geometrical one). The second geometry is from a realistic JET equilibrium scaled up to match the dimensions of ITER. Fig. \ref{fig:equil_geometry} shows the geometry of these 2 equilibria.

\subsection{Testing analytical predictions}

\begin{figure}[ht]
	\centering
	\includegraphics[width=70mm]{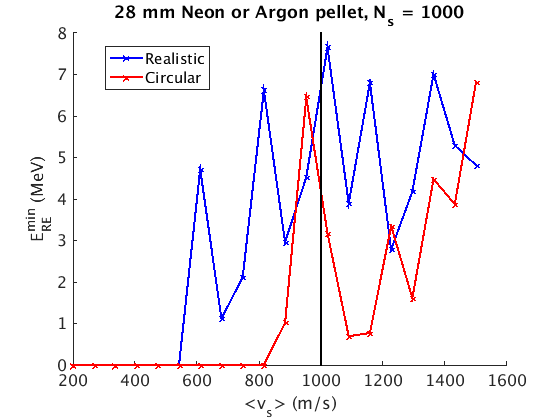}	
	\includegraphics[width=70mm]{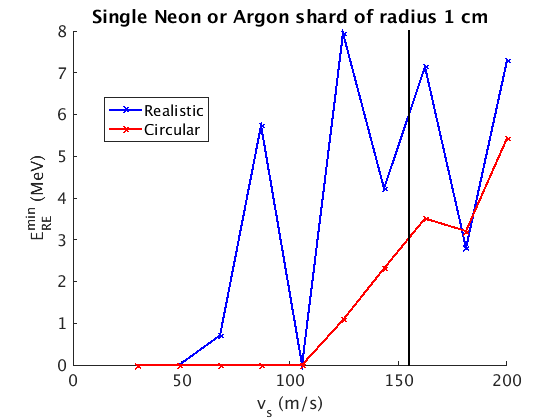}
	\caption{Minimal energy reached by a RE located on the $r_\psi=1$ m surface over the passage of either a cloud of 1000 shards resulting from the shattering of a 28 mm pellet (left) or a single shard of radius 1 cm (right), as a function of the (mean) shard velocity. The material is Neon or Argon in both cases. The vertical lines indicate the critical (mean) shard velocity for stopping REs according to Eq. \ref{eq:t_brake_t_cloud} (left) and \ref{eq:t_stop_t_cloud_single_shard} (right).}
	\label{fig:E_min_vs_vs}
\end{figure}

Fig. \ref{fig:E_min_vs_vs} shows the minimal energy reached by a RE located on the $r_\psi=1$ m surface over the passage of either a cloud of 1000 shards resulting from the shattering of a 28 mm pellet, representative of the `many small shards case' (left), or a single shard of radius 1 cm, representative of the `few large shards' case (right), as a function of the (mean) shard velocity, and for the circular (blue) and realistic (red) equilibrium geometries. The material is Neon or Argon in both cases. Vertical lines indicate the threshold velocity above which REs would not be stopped according to our analytical estimates, see Eq. \ref{eq:t_brake_t_cloud} and \ref{eq:t_stop_t_cloud_single_shard}. It can be seen that the analytical estimate works quite well for the cylindrical geometry, both for `many small shards' and `a few large shards', but overpredicts the critical velocity for the realistic geometry. This is partly a consequence of the fact that in our analytical estimate, we have under-estimated the poloidal perimeter of the flux surface by using $2\pi r_\psi$. In addition, it can be guessed from iso-$\theta^*$ lines in Fig. \ref{fig:equil_geometry} (right) that the $\theta^*$ of a field line after a random number of toroidal turns does not have a homogeneous probability density over $[0,2\pi]$, and in particular has a smaller probability density near $\theta^*=0$ (where shards are located), especially for flux surfaces in the outer part of the plasma. These two effects tend to reduce $E_{loss/turn}$ and thus increase $t_{stop}/t_{pass}$, qualitatively explaining observations.

\subsection{Injection parameters allowing to stop Runaway Electrons}

In order to seek injection parameters allowing to stop REs, we use simulations in realistic geometry with REs initialized at 3 radial positions equally spaced between $\sqrt{\psi_n} = 0.05$ and  $\sqrt{\psi_n} = 0.8$, where $\psi_n \equiv (\psi-\psi_{axis})/(\psi_{LCFS}-\psi_{axis})$ is the normalized poloidal magnetic flux ($\psi_{axis}$ and $\psi_{bnd}$ being the poloidal flux on the magnetic axis and at the last closed flux surface) and at 3 poloidal positions equally spaced in $\theta^*$ between $0$ and $2 \pi$. We do not consider REs beyond $\sqrt{\psi_n} = 0.8$ because as one approaches the LCFS, REs become more and more difficult to stop, probably (mainly) because of the above-mentioned $\theta^*$ effect. RE formation might be avoided in the edge region by applying resonant magnetic perturbations \cite{Papp_PPCF_2011}.

For the case of many small shards and a continuous injection, we consider a shard size corresponding to a 28 mm pellet shattered into 1000 shards. We find that the threshold in terms of solid material volume per unit length (recall the discussion below Eq. \ref{eq:Elt_many_small_1}) to stop REs is equivalent to about 6 (resp. 0.9) 28 mm pellets/meter for Hydrogen (resp. Neon or Argon). These values are close to analytical esimates from Section \ref{many_small_shards}, which may seem surprising since we have seen above that analytical estimates tend to be over-optimistic. This is explained by the fact that in the simulation considered here, the beam of shards extends across the whole plasma, meaning that it intersects each flux surface twice, whereas the analytical estimate takes into account a single intersection.  

For the case of many small shards and repeated discrete injections, we consider the simultaneous injection of a certain number of 28 mm pellets, each shattered into 1000 shards, with an averaged shard velocity $\langle v_s \rangle = 100$ m/s and a relative shard velocity spread $\Delta v_s / \langle v_s \rangle = 0.2$. We find that the threshold number of pellets to stop REs is about 5 (resp. 0.9) for Hydrogen (resp. Neon or Argon).

For the case of a few large shards, we consider a single shard launched from the LFS midplane and travelling across the plasma at $v_s = 100$ m/s. We find that the threshold shard radius for stopping REs is about $2.3$ cm (resp. $1.4$ cm) for Hydrogen (resp. Neon or Argon), which corresponds to about 1.5 (resp. 0.9) times the volume of a 28 mm pellet.

The required injection rates to stop REs found in the different cases above are summarized in Table \ref{table:required_rates}.

\begin{table}
\centering
\begin{tabular}{|c|c|c|c|c|}
  \hline
   & H & Ne & Ar & Unit \\
  \hline
  Many small shards, continuous & 6 & 0.9 & 0.9 &  \# of 28 mm pellets/m \\
  \hline
  Many small shards, repeated & 5 & 0.9 & 0.9 & \# of 28 mm pellets/injection \\
  \hline
  Single shard, repeated & 1.5 & 0.3 & 0.3 & Shard volume converted into \# of 28 mm pellets \\
  \hline
\end{tabular}
  \caption{Required injection rate to stop REs for a(n) (averaged) shard velocity of 100 m/s (see text for details)}
  \label{table:required_rates}
\end{table}

\subsection{Effect of rational surfaces}
\label{sec:rational_q}

\begin{figure}[ht]
	\centering
	\includegraphics[width=70mm]{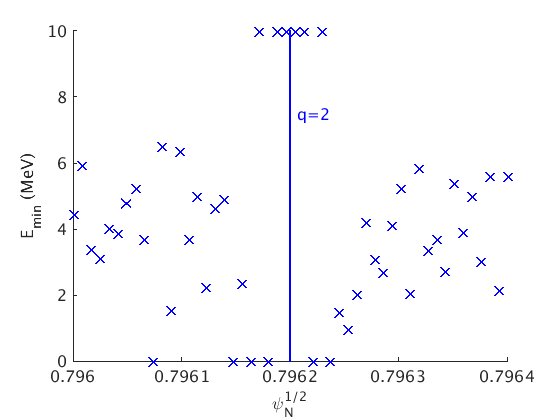}
	\caption{Minimal energy reached by REs during the passage of the shard cloud, as a function of the RE radial position. The vertical line indicates the position of the $q=2$ surface.}
	\label{fig:q_eq_2_resonance}
\end{figure}

It may be expected that some REs located on low order rational surfaces will remain outside the region where shards. While this is clearly true exactly on a rational surface, a more important question is up to what distance from the rational surface this remains true. To address this question, Fig. \ref{fig:q_eq_2_resonance} shows the minimal energy reached by REs during the passage of the shard cloud, as a function of the RE radial position, for a case with 3 large Hydrogen pellets with $\langle v_s \rangle=200$ m/s. REs are initialized away from the region where the cloud passes. The vertical line indicates the position of the $q=2$ surface. It can be seen that indeed, REs initialized around $q=2$ are not braked at all, but this is true only up to a radial distance smaller than $10^{-4}$ of the minor radius. This is because magnetic shear, combined with the fast velocity of REs, causes a strong precession in $\theta^*$ of the REs in a given poloidal plane over the duration of the cloud passage. With an axisymmetric magnetic field, rational surfaces therefore do not seem to pose a major threat to the proposed scheme. On the other hand, the existence of magnetic islands in the plasma could pose a problem since the helical transform is constant inside the islands.

\section{Can shards penetrate?}
\label{sec:penetration}

As is well known from present experiments, a large RE population is capable of vaporizing solid pellets very efficiently. The considered scheme can work only if the RE population always remains sufficiently small that shards can easily travel across the plasma without being vaporized (on the other hand, we assume that liquefaction would not be an issue). The energy deposited per unit volume into a given shard during its travel across the plasma is:

\begin{equation}
	\mathcal{E}_{heat} \simeq \frac{2 G_{av}j_{RE}^{seed}p a}{e v_s}
\end{equation}

where $j_{RE}^{seed}$ is the RE seed current density and $G_{av}$ is the avalanche gain. The `no vaporization' condition can be written $\mathcal{E}_{heat} \ll \mathcal{E}_{vap}$, where $\mathcal{E}_{vap}$ is the shard vaporization heat per unit volume, which can be converted into a condition on $G_{av}$. Since $G_{av} = \exp (\Delta t / t_{e-fold})$ with $t_{e-fold}$ the avalanche e-fold time and $\Delta t$ the time during which the avalanche is free to develop, which in the present case is the temporal gap between (clouds of) shards, one can deduce a condition on $\Delta t$. 

In ITER, the RE seed current from Tritium $\beta$ decay and Compton scattering is expected to be in the Amp\`ere range  \cite{Boozer_PPCF_2019}, such that $j_{RE}^{seed} \sim 0.1$ A/m$^2$. Considering that the poloidal flux variation corresponding to an avalanche e-fold is $\simeq 2.3$ V.s \cite{Boozer_PPCF_2019} and that $V_{loop} = 1.1$ kV, we have $t_{e-fold} \simeq 2.1$ ms. Introducing the value of $\mathcal{E}_{vap}$, which is $\simeq 19$, 120 and 260 MJ/m$^3$ for Hydrogen, Neon and Argon respectively, we estimate that the $\Delta t$ corresponding to vaporization is about 10 ms for all three materials. If we require for example a margin to vaporization by a factor 10, the maximum allowable temporal gap between (clouds of) shards is about 5 ms, corresponding to an injection frequency of 200 Hz. If we multiply the required quantities per injection to stop REs given in Table \ref{table:required_rates} by this frequency, we obtain the injection rates given in Table \ref{table:required_rates_penetration}.

\begin{table}
\centering
\begin{tabular}{|c|c|c|c|}
  \hline
   & H & Ne & Ar  \\
  \hline
  Many small shards, repeated & 1000 & 180 & 180  \\
  \hline
  Single shard, repeated & 300 & 60 & 60  \\
  \hline
\end{tabular}
  \caption{Required injection rate for shards to be able to penetrate, in equivalent number of 28 mm pellets/s (see text for details).}
  \label{table:required_rates_penetration}
\end{table}

The case of a continuous injection of many small shards does not suffer from a requirement related to shard penetration, since RE seeds would be constantly depleted, such that the RE population would always remain far too small to vaporize shards (in addition, REs would not have the possibility to reach large energies). The linear density of pellet volume required to stop REs with this scheme (given in the first row of Table \ref{table:required_rates}) does not depend on $\langle v_s \rangle$, while the injection rate is proportional to $1/\langle v_s \rangle$. If we assume $\langle v_s \rangle = 100$ m/s, we obtained required injection rates equivalent to 600 (resp. 90) 28 mm pellets/s for Hydrogen (resp. Neon or Argon). Comparing these numbers to those in Table \ref{table:required_rates_penetration}, it appears that for a strategy based on the injection of many small shards, a continuous injection requires a roughly twice smaller injection rate than a repeated injection, but required injection rates appear quite large in all cases. With a strategy based on a few large shards, the required injection rate are a bit more moderate.

An important difficulty is that shards take time to reach the center of the plasma. For example, at 100 m/s, they would take about 20 ms, which is much longer than the 5 ms allowable temporal gap found above. This suggests that one cannot afford to wait for the beginning of the CQ to start injections (unless magnetic stochasticity persists and deconfines REs for a long enough period, which is uncertain). A possible solution which would need to be explored is to rely on left-over shards from pre-TQ injections.

\section{Effect of potentially ablated material on Current Quench timescale}
\label{sec:CQ_timescale}

\begin{figure}[ht]
	\centering
	\includegraphics[width=70mm]{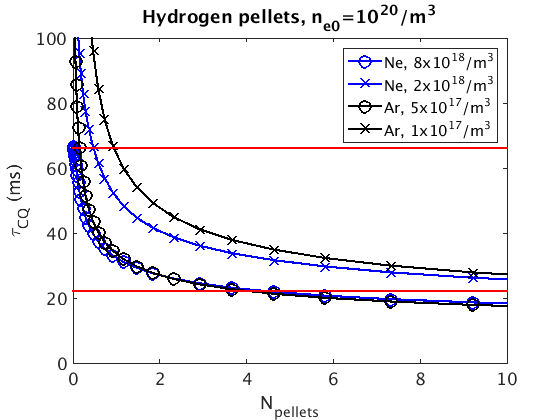}
	\includegraphics[width=70mm]{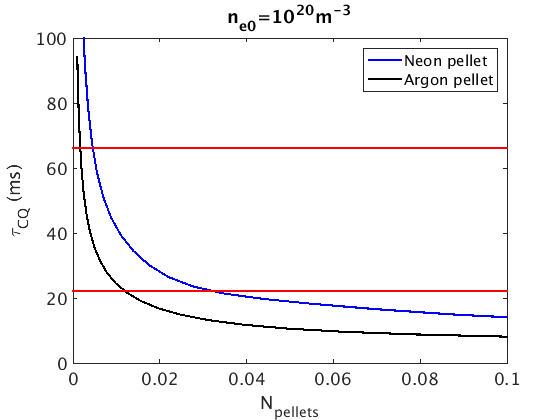}
	\caption{Current Quench exponential timescale as a function of the number of 28 mm pellets assimilated by the plasma, for Hydrogen (left) and Neon or Argon pellets (right). The initial plasma is supposed to have an electron density of $10^{20}$ m$^{-3}$ and, in the case of the Hydrogen pellets, also Neon or Argon impurities whose densities are indicated in the legend.}
	\label{fig:tauCQ_vs_Npellets}
\end{figure}

The last critical aspect to consider to assess the feasibility of the proposed scheme is that if too much of the injected material is assimilated by the plasma, the CQ could become shorter than can be tolerated. Fig. \ref{fig:tauCQ_vs_Npellets} shows how the assimilation of a certain number of Hydrogen (left), Neon or Argon (right) 28 mm pellets is expected to affect the CQ exponential timescale, $\tau_{CQ}$. Red lines indicate the minimal and maximal tolerable values of $\tau_{CQ}$. As in \cite{Martin-Solis_NF_2017}, we estimate the latter as $\tau_{CQ} \simeq La^2/(2R_0\eta)$. We assume a plasma self-inductance $L = 5$ $\mu$H (corresponding to 75 V.s divided by 15 MA), a minor radius $a=2$ m and a major radius $R_0 = 6$ m. The resistivity is calculated as $\eta = 2.8 \times 10^{-8} \times Z_{eff} /T_e^{3/2}$ with $T_e$ in keV and $Z_{eff} = (1+n_{imp}Z_{eff,imp}/n_i)/(1+n_{imp}\langle Z \rangle_{imp}/n_i)$. Here, $\langle Z \rangle_{imp}$ is the averaged charge of the impurity, i.e. $\langle Z \rangle_{imp} \equiv \sum_k Z_{imp,k}n_{imp,k}/n_{imp}$, where the sum is made over charge states of the impurity, and $Z_{eff,imp}$ is the effective charge of the impurity, i.e. $Z_{eff,imp} \equiv \sum_k n_{imp,k}Z_k^2/(\langle Z \rangle_{imp}n_{imp})$. These quantities are functions of $T_e$ obtained from ADAS data \cite{ADAS} assuming coronal equilibrium. The electron temperature is calculated from the power balance between Ohmic heating and radiative losses: $\eta j^2 = n_e n_{imp} L_{rad}$, where $n_e = n_{e0} + \langle Z \rangle_{imp} n_{imp}$, $n_{e0}$ representing the electron density associated to the main ions. We assume $j = 1$ MA/m$^2$. The radiative cooling rate $ L_{rad}$ is a function of $T_e$ calculated from ADAS data assuming coronal equilibrium. 

It appears that for Hydrogen pellets (left plot in Fig. \ref{fig:tauCQ_vs_Npellets}), $\tau_{CQ}$ is not strongly sensitive to the number of assimilated pellets. Typically, up to about 10 pellets could be assimilated while remaining within the tolerable $\tau_{CQ}$  range. A caveat of our model is however that Hydrogen or Deuterium radiation is not included, which could make a key difference and lead to plasma recombination for large amounts of assimilated material, as found in \cite{Vallhagen_2020}. On the other hand, for Neon or Argon pellets, $\tau_{CQ}$ is much more sensitive to the number of assimilated pellets. With the chosen parameters, the maximal tolerable amount of assimilated material is equ
ivalent to about 1\% (resp. 3\%) of a 28 mm pellet for Argon (resp. Neon). 

These tolerable amounts of assimilated material should be compared with estimates of the ablated material. However, ablation rate scalings, like provided in \cite{Sergeev_2006} for example, are not valid at typical CQ electron temperatures \cite{Pegourie_private}. Future work should thus be devoted to modelling pellet ablation during the CQ.


However, even if ablation is sufficiently small, after travelling the plasma, shards would hit the first wall and an important question is whether this could lead to material coming back towards the plasma, either in the form of gas, solid or liquid, and whether this could lead to more material assimilation.

\section{Summary and outlook}
\label{sec:summary}

Table \ref{table:required_rates_summary} summarizes the injection rates required for shards to penetrate and stop REs (combining information from Tables \ref{table:required_rates} and \ref{table:required_rates_penetration}). The unit used in this table in the equivalent number of 28 mm that would need to be injected over a duration of 100 ms, i.e. roughly over the CQ duration. These are clearly large numbers, especially for Hydrogen pellets. For the latter case, required pellet numbers actually appear prohibitive with the presently designed ITER DMS (which cannot inject more than 24 pellets of 28 mm). On the other hand, for Neon or Argon, required numbers may be reachable, although technical aspects need to be investigated. Note that the `many small shards, repeated' case is probably the most relevant one since it is closest to what would be obtained with repeated SPI.

\begin{table}
\centering
\begin{tabular}{|c|c|c|c|}
  \hline
   & H & Ne & Ar  \\
   \hline
  Many small shards, continuous & 60 & 9 & 9  \\
  \hline
  Many small shards, repeated & 100 & 18 & 18  \\
  \hline
  Single shard, repeated & 30 & 6 & 6  \\
  \hline
\end{tabular}
  \caption{Required injection rate for shards to penetrate and stop REs, in equivalent number of 28 mm pellets over 100 ms (see text for details).}
  \label{table:required_rates_summary}
\end{table}

Table \ref{table:max_pellets_CQ_duration} summarizes how many 28 mm pellets could be assimilated before the CQ becomes too short (see Fig. \ref{fig:tauCQ_vs_Npellets}).

\begin{table}
\centering
\begin{tabular}{|c|c|c|c|}
  \hline
   & H & Ne & Ar  \\
   \hline
  Max. \# of assimilated 28 mm pellets for acceptable $\tau_{CQ}$ & 5-10 & 0.03 & 0.01  \\
  \hline
\end{tabular}
  \caption{Maximum number of 28 mm pellets which could be assimilated before the CQ becomes too short.}
  \label{table:max_pellets_CQ_duration}
\end{table}

For Neon or Argon, there are more than two orders of magnitude between numbers in Tables \ref{table:required_rates_summary} and \ref{table:max_pellets_CQ_duration}, meaning that the assimilated fraction would have to be extremely small. The development of pellet ablation models in CQ plasmas, as well as experiments on SPI during the CQ in present machines, would help assess whether this could be realistic.

In the present study, we have considered only Hydrogen, Neon or Argon as candidate materials in order to match the capabilities of the presently designed ITER DMS. However, other materials with ideally a larger stopping power, lower radiation rate at typical CQ temperatures and lower ablation rate, would be better suited for the proposed scheme. Some effort should be devoted to looking for suitable materials.

Another aspect which would need to be looked into is the effect of the vertical plasma motion during the CQ. One should ensure that shards do not `miss their target' as a result of this motion. An injection scheme ensuring a wide angular dispersion of the shards would be helfpul in this respect.






\end{document}